\documentclass[12pt,review]{iopart}
\usepackage{iopams,amsmath2}
\usepackage{graphicx,color}

\newcommand{\tn}{\textnormal}
\newcommand{\cprb}[3]{Phys.~Rev.~B {\bf #1}, #2 (#3)}
\newcommand{\cprl}[3]{Phys.~Rev.~Lett.~{\bf #1}, #2 (#3)}

\newcommand{\cjpa}[3]{J.~Phys.~A {\bf #1}, #2 (#3)}

\newcommand{\cjpamg}[3]{J.~Phys.~A: Math.~Gen {\bf #1}, #2 (#3)}
\newcommand{\cpre}[3]{Phys.~Rev.~E {\bf #1}, #2 (#3)}

\definecolor{darkred}{rgb}{0.90,0,0}
\definecolor{darkgreen}{rgb}{0,0.60,.2}
\definecolor{darkblue}{rgb}{0,0,1}
\definecolor{pink}{rgb}{1,0,1}
\definecolor{grey}{cmyk}{0,0,0,0.25}
\definecolor{orange}{cmyk}{0,0.6,0.8,0}

\newcommand{\onlinecite}[1]{\cite{#1}}

\begin{document}
\title[Reducing the numerical effort of finite-temperature DMRG]{Reducing the numerical effort of finite-temperature density matrix renormalization group transport calculations}

\author{C.\ Karrasch, J.\ H.\ Bardarson, and J.\ E.\ Moore}

\address{Department of Physics, University of California, Berkeley, California 95720, USA}

\address{Materials Sciences Division, Lawrence Berkeley National Laboratory, Berkeley, CA 94720, USA}

\begin{abstract}
Finite-temperature transport properties of one-dimensional systems can be studied using the time dependent density matrix renormalization group via the introduction of auxiliary degrees of freedom which purify the thermal statistical operator. We demonstrate how the numerical effort of such calculations is reduced when the physical time evolution is augmented by an additional time evolution within the auxiliary Hilbert space. Specifically, we explore a variety of integrable and non-integrable, gapless and gapped models at temperatures ranging from $T=\infty$ down to $T/\tn{bandwidth}=0.05$ and study both (i) linear response where (heat and charge) transport coefficients are determined by the current-current correlation function and (ii) non-equilibrium driven by arbitrary large temperature gradients. The modified DMRG algorithm removes an `artificial' build-up of entanglement between the auxiliary and physical degrees of freedom. Thus, longer time scales can be reached.
\end{abstract}

\maketitle


\section{Introduction}
\label{sec:intro}

A physical system is usually characterized by its response to perturbations. In transport setups, one studies charge or energy currents driven by voltage or temperature gradients. From the theoretical point of view this is complicated -- computing the quantum mechanical time evolution of a system in non-equilibrium is one of the most active areas of research in condensed matter physics. When the external perturbations are small, one can simplify the problem by resorting to the Kubo formalism. E.g., the charge conductivity $\sigma(\omega)$ describes the \textit{linear response} current $J$ induced by a \textit{small} electric field (we will be more specific below):
\begin{equation}
\sigma(\omega) \sim   \int e^{i\omega t} \langle J(t)J\rangle dt~,
\end{equation}
where the dynamical correlation function $\langle J(t)J\rangle$ is calculated in \textit{thermal equilibrium}:
\begin{equation}\label{eq:intro0}
\langle A(t) B \rangle = \tn{Tr} \left( \rho_T\, e^{iHt}Ae^{-iHt}B \right)~,~~\rho_T = \frac{e^{-H/T}}{\tn{Tr}\,e^{-H/T}}~,
\end{equation}
with $H$ being the Hamiltonian of the system and $T$ denoting the temperature. Unfortunately, computing transport coefficients such as $\sigma(\omega)$ is generally still difficult: Even if one knows the exact thermal density matrix $\rho_T$ (or the exact ground state), extracting correlation functions, which couple all excitations, remains a formidable task.

The key question posed and addressed in this work is: How can linear-response and non-equilibrium transport properties of one-dimensional (1d) systems at finite temperature be calculated \textit{efficiently} using the density matrix renormalization group (DMRG) \cite{dmrgrev0,dmrgrev}? DMRG was originally devised \cite{white1,white2} as tool to accurately determine \textit{ground states} of 1d Hamiltonians. The reason for its success became understandable when it was formulated using matrix product states (MPS) \cite{mps1,mps2,verstraete1,verstraete2,arealaw2}: The area law \cite{arealaw} stipulates that the ground states of 1d systems governed by local Hamiltonians are only entangled locally; this implies that they can be expressed efficiently using a MPS with a small bond dimension $\chi$ (which encodes the amount of entanglement). The MPS describing a given system can be determined variationally -- this is the very core of a ground state DMRG calculation \cite{white1,white2}. One way to compute \textit{correlation functions} $\langle A(t)B\rangle$ at $T=0$ is to directly simulate the time evolution (for other approaches see Refs.~\onlinecite{dmrgrev0,dyndmrg0,dyndmrg0a,dyndmrg,jeckelmann,jan,piet})
\begin{equation}\label{eq:intro0b}
e^{-iHt}B|\tn{ground state}\rangle
\end{equation}
using a time-dependent DMRG framework \cite{tdmrg1,tdmrg2,tdmrg3,tdmrg3a,tdmrg4,tebd,noneqprosen1,verstraete0,fleischhauer}. The corresponding algorithm can again be formulated elegantly using matrix product states. The physical growth of entanglement implies that the bond dimension needed to approximate $\langle A(t)B\rangle$ to a certain accuracy grows with time. This limits the accessible time scales.

Standard DMRG methods allow computing the time evolution of a \textit{pure} state and are thus not directly applicable at $T>0$. Various approaches for simulating finite-temperature dynamics using DMRG have been explored within the literature \cite{verstraete,vidalop,tmrg0,tmrg0a,metts,trick2a,trick2b,tmrg1,tmrg2,sirker1,sirker2,dmrgT,barthel,fiete,drudepaper}. These include probabilistic sampling over an appropriately chosen set of pure states \cite{metts}, schemes which time-evolve operators instead of states \cite{trick2a,trick2b}, transfer-matrix DMRG \cite{tmrg1,tmrg2,sirker1,sirker2}, and exact representations through purification \cite{dmrgT,barthel,fiete}. Purification expresses the thermal statistical operator $\rho_T$ as a partial trace over a pure state $|\Psi_T\rangle$ living in an enlarged Hilbert space where auxiliary degrees of freedom $Q$ encode the thermal bath \cite{purification}:
\begin{equation}
\rho_T = \tn{Tr}_Q |\Psi_T\rangle\langle\Psi_T|~.
\end{equation}
One of the main advantages of this approach is that all the standard methods for time evolving quantum states within DMRG are directly applicable. In this work, we employ the time evolving block decimation \cite{tdmrg1,tebd}.

One of the first applications of finite-temperature `purification DMRG' to dynamical problems was the calculation of spin-spin correlation functions of integrable spin-$1/2$ Heisenberg chains \cite{barthel}. While DMRG yields data which are `numerically exact' (this is verified by comparing with analytic results available for `non-interacting' models \cite{tmrg1,tmrg2,barthel}), the time scales accessible at finite temperatures are considerably smaller than at $T=0$. This observation, which might be one reason why studies of finite-$T$ dynamics \cite{trick2a,trick2b,tmrg1,tmrg2,sirker1,sirker2,barthel,fiete,drudepaper,integrabilitypaper,thermalpaper,drudepaper2} are rarer than their $T=0$ counterparts, can be understood as follows. Assume we want to compute a ground state correlation function, i.e.~evaluate Eq.~(\ref{eq:intro0b}). Under the time evolution, the entanglement grows \textit{locally} around the region on which $B$ acted. In contrast, at $T>0$ where one needs to calculate (see below for details)
\begin{equation}\label{eq:intro1}
e^{-iHt}B|\Psi_T\rangle~, 
\end{equation}
the entanglement generally increases \textit{homogeneously} throughout the whole system. This holds true even for $B=1$, i.e.~when the state $|\Psi_T\rangle$ is exposed to a supposedly trivial time evolution \cite{drudepaper}. The reason for this is that purification is not unique and various representations of the same $\rho_T$ differ in their degree of entanglement between the auxiliary and physical degrees of freedom. Even when one starts out with a purification that minimizes this entanglement, it rapidly grows under the DMRG time evolution. This observation naturally leads to the question: Can that very same `freedom' be used to our advantage to undo the non-physical growth of entanglement? In other words, is there a unitary transformation $U_Q(t)$ acting on the auxiliary Hilbert space $Q$ which removes the artificial entanglement? In Ref.~\onlinecite{drudepaper} we proposed to time-evolve $Q$ backwards in time with the physical Hamiltonian acting on the auxiliary degrees of freedom:
\begin{equation}
U_Q(t) = e^{+i\tilde H t}~.
\end{equation}
This renders the time evolution of $|\Psi_T\rangle$ trivial, and the evaluation of Eq.~(\ref{eq:intro1}) is therefore eventually only plagued by an entanglement building up around the region where $B$ acts in complete analogy to the ground state calculation (the physical reason being quasi-locality \cite{trick2a,trick2b}). This generically leads to a slower increase of the bond dimension $\chi$, and thus longer time scales can be reached.

In Ref.~\onlinecite{drudepaper}, the potential of the modified DMRG algorithm was demonstrated for the spin-spin correlation function $\langle S^z_n(t)S^z_m\rangle$ of the XXZ chain at zero anisotropy $\Delta=0$, which maps to free fermions and thus allows for an exact (benchmark) solution. A more thorough comparison between the numerical effort of the standard \cite{barthel} and modified \cite{drudepaper} algorithms in calculating $\langle S^z_n(t)S^z_m\rangle$ for arbitrary $\Delta$ can be found in Ref.~\onlinecite{trick2b}, where further optimizations using operator-space DMRG were explored. One of the ideas of Ref.~\onlinecite{trick2b} is to use time translation invariance to rewrite Eq.~(\ref{eq:intro0}) as
\begin{equation}\label{eq:intro2}
\langle A(t)B\rangle = \langle A(t/2)B(-t/2)\rangle~,
\end{equation}
which allows accessing times \textit{twice as large} without additional effort. Refs.~\onlinecite{trick2a,trick2b} also show that for certain scenarios it can be more efficient to redistribute the evaluation of $\exp(-iHt)$ and $\exp(-\beta H)$ over two DMRG simulations.

From the point of view of physical applications, the modified DMRG schemes were used to investigate current correlation functions and the Drude weight of the integrable XXZ chain at intermediate to high temperatures \cite{drudepaper,drudepaper2}, scaling properties of the DC conductivity in presence of non-integrable perturbations \cite{integrabilitypaper}, non-equilibrium induced by temperature gradients \cite{thermalpaper}, and spectral functions of hardcore bosons \cite{trick2a}. For reasons of completeness, we mention other approaches to linear response transport properties of the XXZ model (and related ones). These include exact Bethe ansatz calculations \cite{bethe1,bethe2,bethe3,bethe4}, integrability arguments \cite{integrability1,integrabiliy2,prosen,znidaric1,znidaric2,znidaric3}, field theories \cite{sirker1,sirker2,lowenergy1,lowenergy2,lowenergy3}, quantum Monte Carlo \cite{qmc1,qmc2,qmc3}, exact diagonalization \cite{ed1,ed2,ed3,ed4,ed5}, transfer matrix DMRG \cite{sirker1,sirker2}, and dynamical DMRG \cite{jeckelmann}. For a non-exhaustive list of prior works on non-equilibrium thermal transport see Refs.~\onlinecite{noneqprosen1,noneqtherm1,noneqtherm2,fabiannoneq,fabiannoneq2,prosen2,noneqstein,prosentherm,free2,bruneau,doyon,doyon2}. 

The first and foremost goal of this paper is to present extensive quantitative data on how the build-up of entanglement is reduced if the auxiliary Hilbert space $Q$ is time-evolved with the physical Hamiltonian but reversed time. We focus on transport properties in linear response \cite{drudepaper,integrabilitypaper,drudepaper2} and in thermal non-equilibrium \cite{thermalpaper}. Specifically, we study (i) homogeneous gapless and gapped spin-$1/2$ XXZ chains, also in presence of various perturbations which break integrability, (ii) the quantum Ising model, and (iii) impurity setups of a quantum dot connected to non-interacting leads. For temperatures from $T=\infty$ down to $T/\tn{bandwidth}=0.05$, the modified DMRG algorithm leads to a slower increase of the MPS dimension $\chi$. Only at low $T$ does the standard approach $U_Q=1$ become more efficient (see Ref.~\onlinecite{trick2b} for details). For most (but not all; we will be more specific below) problems studied in this work, $T/\tn{bandwidth}=0.05$ is a low enough temperature to correspond to the $T=0$ limit, which one can establish, e.g., by comparing with field theory results \cite{doyon}. We show the typical behavior of $\chi$ on the relevant physical time scales for each problem at hand (e.g., on the time scale at which non-equilibrium currents generically reach their steady state). We reiterate how Eq.~(\ref{eq:intro2}) can be implemented within the purification approach and illustrate (following Refs.~\onlinecite{trick2a,trick2b}) how it allows accessing times twice as large.

As the second purpose of this paper, we present technical details of the implementation of the algorithm. E.g., we show how to time-evolve next-nearest neighbors (closely following Ref.~\onlinecite{dmrgrev}), which is necessary because physical degrees of freedom are separated by an auxiliary site due to the purification. We discuss the numerical accuracy of our data, compare with exact results, investigate to what extend the choice of $U_Q(t)=\exp(i\tilde Ht)$ is optimal \cite{trick2a,trick2b}, and provide further evidence for the reliability of linear prediction extrapolation schemes \cite{barthel,integrabilitypaper,linpred,linpred2} for the spin-spin correlation function of the isotropic Heisenberg chain.

\section{Models and methods}
\label{sec:method}

\subsection{Models}

As a prototypical model, we consider a chain of interacting spin-$1/2$ degrees of freedom $S^{x,y,z}_n$ governed by local Hamiltonians
\begin{equation}\label{eq:h}
h_{n}  =  J_n \big(S^x_{n}S^x_{n+1} + S^y_{n}S^y_{n+1} + \Delta_n S^z_{n}S^z_{n+1}\big)  + b_n (S_n^z-S_{n+1}^z) ~,
\end{equation}
or equivalently spinless fermions through a Jordan-Wigner transformation. By choosing the couplings $J_n$, $\Delta_n$, and $b_n$ appropriately:
\begin{equation}\label{eq:parahom}
J_n =
\begin{cases}
1 & n \tn{ odd} \\
\lambda & n \tn{ even}    
\end{cases}~,~~
\Delta_n = \Delta~,~~ b_n= \frac{(-1)^n b}{2}~~,
\end{equation}
we can study systems which are gapless or gapped, and investigate the role of integrability. For $\lambda=1$ and $b=0$, Eq.~(\ref{eq:h}) can be diagonalized via Bethe ansatz \cite{bethegs}; the model is non-integrable otherwise. The energy spectrum is gapless for $|\Delta|\leq1$ and gapped for $\Delta>1$. A gap opens for $\lambda<\lambda_c$ or $b>b_c$, where $\lambda_c<1$ and $b_c>0$ if $-1<\Delta<-1/\sqrt{2}$ \cite{sato,stagfield,integrabilitypaper}. In addition, we study the quantum Ising model
\begin{equation}\label{eq:h2}
h_n = - 4 S^z_{n}S^z_{n+1} -  g( S_n^x+S_{n+1}^x )~.
\end{equation}

\subsection{Transport properties}

\subsubsection{Linear response.}

We first consider a homogenous system of size $L$ governed by $H = \sum_{n=1}^{L-1} h_n$ within linear response. The optical charge (C) and energy (E) conductivities can be computed via the Kubo formula
\begin{equation}\label{eq:cond}
\sigma(\omega) = \frac{1}{\omega L}\int_0^\infty e^{i\omega t} \langle [J(t),J]\rangle dt
 = \frac{1-e^{-\omega/T}}{\omega L} \int_0^\infty e^{i\omega t} \langle J(t)J\rangle dt~,
\end{equation}
where the corresponding current operators $J=\sum_n j_n$ are defined through a continuity equation \cite{commentconteq}:
\begin{equation}\label{eq:current}\begin{split}
\partial_t h_n = j_{\textnormal{E},n}-j_{\textnormal{E},n+1}& ~\Rightarrow~  J_\textnormal{E} = i \sum_{n=2}^{L-1} [h_{n-1},h_n]~,\\
\partial_t S_n^z = j_{\textnormal{C},n}-j_{\textnormal{C},n+1}& ~\Rightarrow~  J_\textnormal{C} = i \sum_{n=2}^{L-1} [h_{n-1},S_n^z]~.
\end{split}\end{equation}
One usually decomposes the real part of $\sigma(\omega)$ as 
\begin{equation}\label{eq:cond2}
\tn{Re }\sigma(\omega) = 2\pi D \delta(\omega) + \sigma_{\tn{reg}}(\omega)~,
\end{equation}
where the so-called Drude weight $D$ is the prefactor of the singular contribution. $D$ can be determined from the long-time asymptote of the current-current correlation function
\begin{equation}\label{eq:drude}
D = \lim_{t\to\infty}\lim_{L\to\infty} \frac {\textnormal{Re } \langle J(t)J\rangle}{2LT}~.
\end{equation}
As already stated above, a time-dependent equilibrium correlation function is defined as
\begin{equation}\label{eq:corr}
\langle A(t) B \rangle = \tn{Tr} \left( \rho_T\, e^{iHt}Ae^{-iHt}B \right)~,~~\rho_T = \frac{e^{-H/T}}{Z_T}~,
\end{equation}
with $Z_T=\tn{Tr}\exp(-H/T)$ denoting the partition function.

\begin{figure}[t]
\includegraphics[height=0.53cm,clip]{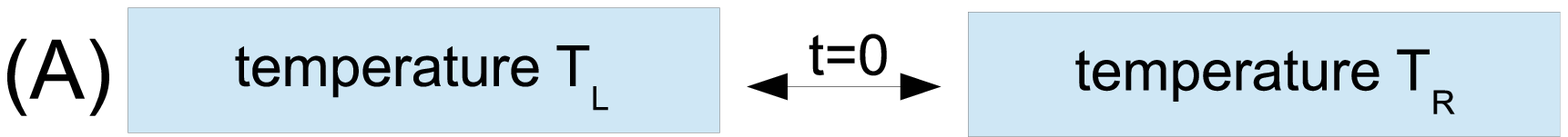}\hspace*{0.5cm} 
\includegraphics[height=0.53cm,clip]{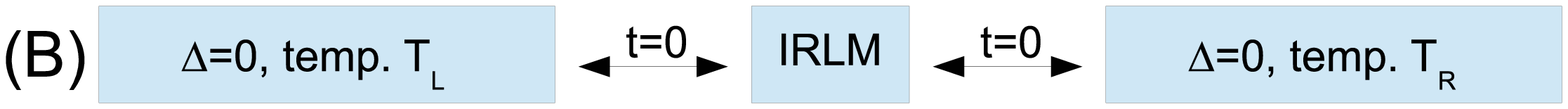} 
\caption{(Color online) The non-equilibrium setups studied in this work. A: Two interacting chains of length $L/2$ which are initially in thermal equilibrium at temperatures $T_L$ and $T_R$ are coupled at time $t=0$. B: Two non-interacting chains are coupled at time $t=0$ via an interacting resonant level model (IRLM).}
\label{fig:setup}
\end{figure}

\subsubsection{Non-equilibrium.}
In addition to linear response, we study two thermal non-equilibrium setups. The first [labeled `non-equilibrium A' and depicted in Figure \ref{fig:setup}(A)] is introduced via the following protocol: We initially consider two separate chains,
\begin{equation}
H_0 = H_L + H_R = \sum_{n=1}^{L/2-1} h_n + \sum_{n=L/2+1}^{L-1} h_n ~,
\end{equation}
each being in thermal (grand-canonical) equilibrium at temperatures $T_L$ and $T_R$. The corresponding density matrix factorizes,
\begin{equation}
\rho_0=\rho_L\otimes\rho_R~,~~ \rho_i=\frac{\exp(-H_i/T_i)}{\tn{Tr}\exp(-H_i/T_i)}~,~~i=L,R~.
\end{equation}
At time $t=0$, the chains are coupled through $h_{L/2}$, and the time evolution of any observable $A$ is computed using $H=H_0+h_{L/2}$: 
\begin{equation}\label{eq:rhot}
\langle A(t)\rangle = \tn{Tr}\, \rho(t) A~,~~  \rho(t) = e^{iHt}\rho_0e^{-iHt}~.
\end{equation}
In the second setup [`non-equilibrium B'; see Figure \ref{fig:setup}(B)] we investigate two non-interacting chains $\Delta=b=0$, $\lambda=1$ of length $L/2$,
\begin{equation}
H_0 = H_L + H_R = \sum_{n=1}^{L/2-1} h_n + \sum_{n=L/2+2}^{L} h_n~, 
\end{equation}
at different temperatures $T_L$ and $T_R$. At $t=0$, they are coupled via an interacting resonant level model:
\begin{equation}\label{eq:irlm}\begin{split}
& h_\tn{IRLM} =
 t' \big(S^x_{L/2}S^x_{L/2+1} + S^y_{L/2}S^y_{L/2+1} + U S^z_{L/2}S^z_{L/2+1}\big) \\
& +  t' \big(S^x_{L/2+1}S^x_{L/2+2} + S^y_{L/2+1}S^y_{L/2+2} + U S^z_{L/2+1}S^z_{L/2+2}\big)~.
\end{split}\end{equation}
The site $n=L/2+1$ is initially in an equal superposition of up and down states (i.e., formally at infinite temperature).

\subsection{DMRG}
\label{sec:aux}

In this Section, we give a brief overview of the DMRG method \cite{dmrgrev0,dmrgrev,white1,white2}. More details can be found in the Appendix. In order to evaluate Eqs.~(\ref{eq:corr}) or (\ref{eq:rhot}) by a standard DMRG algorithm (which time evolves wave functions) one first needs to purify the thermal density matrix $\rho_T$ by introducing an auxiliary Hilbert space $Q$ such that $\rho_T = \tn{Tr\,}_{Q} |\Psi_T\rangle\langle\Psi_T|$. This is analytically possible only at $T=\infty$ where $\rho_T$ factorizes. However, $|\Psi_T\rangle$ can be obtained from $|\Psi_\infty\rangle$ by applying an imaginary time evolution, $|\Psi_T\rangle=e^{-H/(2T)}|\Psi_\infty\rangle$ \cite{verstraete,dmrgT,barthel}. The correlation function of Eq.~(\ref{eq:corr}) is \textit{exactly} recast as
\begin{equation}\label{eq:corr2B}
\langle A(t)B\rangle = \frac{\langle \Psi_\infty| e^{-\frac{H}{2T}}e^{iHt} A e^{-iHt} B e^{-\frac{H}{2T}}|\Psi_\infty\rangle}{\langle\Psi_\infty| e^{-\frac{H}{T}}|\Psi_\infty\rangle}~,
\end{equation}
and this object is directly accessible in standard time-dependent DMRG frameworks \cite{tdmrg1,tdmrg2,tdmrg3,tdmrg3a,tdmrg4,tebd}. It is convenient to first express the initial state $|\Psi_\infty\rangle$ in terms of a matrix product state \cite{mps1,mps2,verstraete1,verstraete2},
\begin{equation}
|\Psi_\infty\rangle = \sum_{\sigma_n,\sigma_{n_Q}} A^{\sigma_1}A^{\sigma_{1_Q}}\cdots A^{\sigma_{L}}A^{\sigma_{L_Q}}|\sigma_1\sigma_{1_Q}\ldots\sigma_{L}\sigma_{L_Q}\rangle~,
\end{equation}
where
\begin{equation}\label{eq:mpsB}
A^{\sigma_i}_{a_ia_{i+1}} = \Lambda_{a_i}^i\Gamma_{a_ia_{i+1}}^{\sigma_i}~.
\end{equation}
Here and in the following $\sigma_i$ is a short hand for either a physical or auxiliary degrees of freedom: $\sigma_i\in\{\sigma_n,\sigma_{n_Q}\}$. The initial matrices associated with $|\Psi_\infty\rangle$ read
\begin{align}
\Gamma^{\sigma_{n_{\phantom{Q}}}=\uparrow}& = (1~~0) &
\Gamma^{\sigma_{n_{\phantom{Q}}}=\downarrow}& = (0~~-1) \nonumber \\
\Gamma^{\sigma_{n_Q}=\uparrow}& = (0~~1/\sqrt{2})^T &
\Gamma^{\sigma_{n_Q}=\downarrow}& = (1/\sqrt{2}~~0)^T ~,
\end{align}
as well as $\Lambda^i=1$. After factorizing the evolution operators $\exp(-\lambda H)$ using a second or fourth order Trotter decomposition, they can be successively applied to Eq.~(\ref{eq:mpsB}). At each time step $\Delta\lambda$, two singular value decompositions are carried out to update three consecutive matrices. The matrix dimension $\chi$ is dynamically increased such that at each time step the sum of all squared discarded singular values is kept below a threshold value $\epsilon$.

An exact modification to the finite-temperature DMRG algorithm (which allows accessing longer time scales) was recently introduced in Ref.~\onlinecite{drudepaper}: One has the analytic freedom to apply any time-dependent unitary transformation
\begin{equation}
U_Q(t) = \sum_{\sigma_{1_Q}\ldots \sigma_{L_Q}\atop\sigma_{1_Q}'\ldots \sigma_{L_Q}'}
C_{\sigma_{1_Q}'\ldots \sigma_{L_Q}'}^{\sigma_{1_Q}\ldots \sigma_{L_Q}}(t) |\sigma_{1_Q}\ldots \sigma_{L_Q}\rangle\langle \sigma_{1_Q}'\ldots \sigma_{L_Q}'|
\end{equation}
to the in principle inert auxiliary sites $\sigma_{n_Q}$ -- physical quantities are determined by the trace over $Q$ and are thus not affected by $U_Q(t)$:
\begin{equation}
[U_Q(t), \sigma_n] = 0~.
\end{equation}
For example, Eq.~(\ref{eq:corr2B}) can be rewritten as
\begin{equation}\label{eq:mporef}
\langle \Psi_T|e^{iHt} A e^{-iHt} e^{iHt'} B e^{iHt'} |\Psi_T\rangle
= \langle \Psi_T| U^\dagger (t) A U(t) U^\dagger(t') B U(t')|\Psi_T\rangle~,
\end{equation}
where 
\begin{equation}
U(t) = e^{-iHt} U_Q(t)~.
\end{equation}
Put differently, purification is not unique. It turned out \cite{trick2a,trick2b,drudepaper,integrabilitypaper,thermalpaper} that choosing
\begin{equation}
U_Q(t) = e^{+i\tilde Ht}~,~~\tilde H = H\big(\sigma_n\to \sigma_{n_Q}, \tn{magnetic fields reversed}\big)~,
\end{equation}
i.e., time-evolving the auxiliary sites with the physical Hamiltonian but reversed time, leads to a slower increase of $\chi$ and thus longer time scales can be reached. An intuitive way of understanding this will be given below. Only at low temperatures does using $U_Q=1$ become more efficient \cite{trick2a,trick2b}. In practice it is most convenient to implement
\begin{equation}
e^{-iHt}e^{i\tilde Ht} = e^{-iH\Delta t}e^{i\tilde H\Delta t}e^{-iH\Delta t}e^{i\tilde H\Delta t}\ldots~~.
\end{equation}

\section{Results}
\label{sec:results}

In this section we present our main results. We first quantify how the entanglement growth is reduced by time-evolving the auxiliaries with the physical Hamiltonian but reversed time. In an extension of earlier studies (see Refs.~\onlinecite{trick2a,trick2b,drudepaper} and the discussion in Sec.~\ref{sec:intro}), we study gapless and gapped XXZ chains in presence of non-integrable perturbations and focus on transport problems both in linear response and out of equilibrium. We investigate to what extend $U_\tn{aux}(t)=\exp(it\tilde H)$ is optimal within our approach. In passing, we provide another example of the stability of linear prediction (see also Refs.~\onlinecite{dmrgrev,barthel,integrabilitypaper,linpred,linpred2}) and recast (one of) the ideas of Refs.~\onlinecite{trick2a,trick2b} in the language of purification.

\begin{figure}[t]
\centering\includegraphics[width=0.5\linewidth,clip]{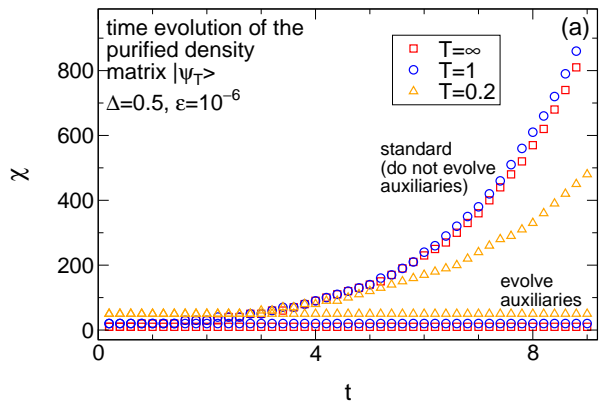}
\caption{(Color online) MPS dimension during the time evolution of the state $|\Psi_T\rangle$ which purifies the thermal density matrix of a XXZ chain. }
\label{fig:chi0}
\end{figure}

\subsection{Prelude: Time evolution of $|\Psi_T\rangle$}
It is instructive to first consider the trivial case of $A=B=1$ in Eq.~(\ref{eq:corr2B}), i.e.
\begin{equation}
1 =\frac{\langle\Psi_T|e^{iHt} e^{-iHt}| \Psi_T\rangle}{\langle\Psi_T|\Psi_T\rangle}
= \frac{\langle\Psi_T|e^{iHt}U_Q^\dagger(t) e^{-iHt}U_Q(t)|\Psi_T\rangle}{\langle\Psi_T|\Psi_T\rangle}~.
\end{equation}
Figure \ref{fig:chi0} shows the evolution of the bond dimension $\chi$ during the calculation of $e^{-iHt}U_Q(t)|\Psi_T\rangle$. For the standard approach $U_Q(t)=1$, $\chi$ increases with time: the state $e^{-iHt}|\Psi_T\rangle$ which purifies the density matrix becomes successively more entangled. Choosing $U_Q(t)=e^{i\tilde Ht}$ removes this artifact. This immediately indicates why longer time scales can be reached in DMRG evaluations of, e.g., current correlation functions $e^{-iHt}U_Q(t)j_{n}|\Psi_T\rangle$. The entanglement only builds up locally around site $n$ (the physical reason being quasi-locality \cite{trick2a,trick2b}), and likewise for our non-equilibrium setup it grows locally around the interface region over which the initial temperature gradient falls off. The artificial global build-up of entanglement is removed if the auxiliaries are evolved in time. Only at low $T$ choosing $U_Q=1$ becomes more efficient \cite{trick2a,trick2b} since the time evolution of the ground state is trivial (the latter is indicated in Figure \ref{fig:chi0}: $\chi$ grows more slowly at low $T=0.2$ than it does at high $T$).

An intuitive understanding of why the particular choice of $U_Q(t) = e^{+i\tilde Ht}$ removes the `artifical' entanglement growth was recently provided in Refs.~\onlinecite{trick2a,trick2b}: In an operator-space formulation, the modified DMRG algorithm corresponds to a Heisenberg time evolution of the matrix product operators representing $A$ and $B$ in Eq.~(\ref{eq:mporef}).  If $A$ (and likewise for $B$) is local, most terms in the first Trotter step $e^{iH\Delta t}Ae^{-iH\Delta t}$ cancel out. As time evolves,
\begin{equation}
\ldots e^{iH\Delta t}e^{iH\Delta t}Ae^{-iH\Delta t}e^{-iH\Delta t}\ldots~~,
\end{equation}
entanglement can only build up gradually around the region on which $A$ acted due to the so-called light-cone effect in nonrelativistic systems: Lieb-Robinson bounds \cite{liebrobinson} generically stipulate that correlation functions $\langle A(t)B\rangle$ of local operators $A$ and $B$ fall off exponentially for $x>vt$, with $x$ denoting the spatial distance between the regions on which $A$ and $B$ act, and $v$ being some velocity with which excitations propagate.

\begin{figure}[t]
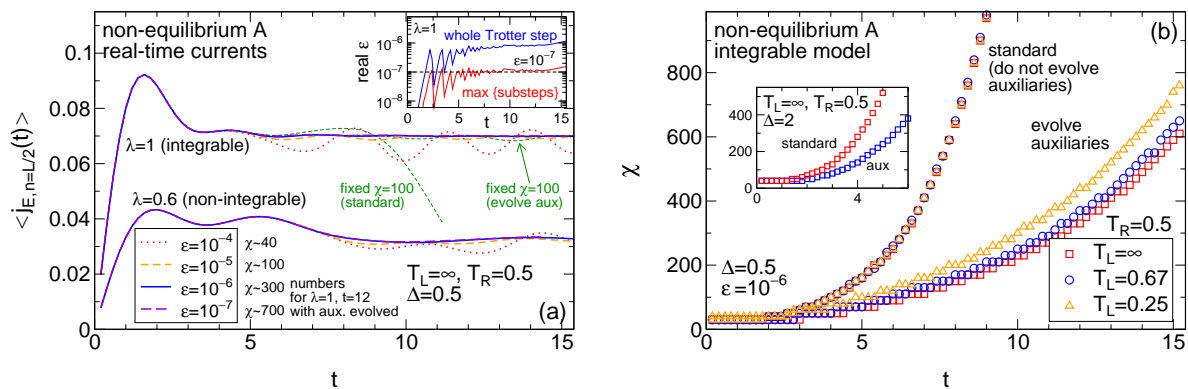

\includegraphics[width=0.48\linewidth,clip]{current.eps} \hspace*{0.02\linewidth}
\includegraphics[width=0.48\linewidth,clip]{chi2.eps}
\caption{(Color online) DMRG time evolution of an XXZ chain of length $L=200$ featuring an initial sharp temperature gradient $T_L\neq T_R$. The system is gapless for $z$-anisotropies $|\Delta|\leq1$ and gapped otherwise; it becomes non-integrable in presence of a finite dimerization $\lambda<1$. Note that for $\Delta=0.5$ and $\lambda=1$ asymptotic low-temperature behavior described by a field theory \cite{doyon} sets in for $T\lesssim0.2$. (a) Energy current at the interface. Choosing a discarded weight control parameter around $\epsilon\sim10^{-6}$ at a Trotter stepsize of $\Delta t=0.2$ is sufficient to accurately obtain its steady-state value. Inset: The \textit{actual} discarded weights during a whole Trotter step and during one substep for $\Delta\chi=20$. The difference to $\epsilon$ is explained in the main text. (b) Evolution of the dimension of the corresponding matrix product state. If the auxiliary degrees of freedom which purify the thermal density matrix are time-evolved with the physical Hamiltonian but reversed time (which is an exact modification to the standard algorithm), the build-up of entanglement is reduced. The computational cost of a singular value decomposition (which dominates the DMRG algorithm) scales as $\chi^3$.  }
\label{fig:noneq1}
\end{figure}

\subsection{Reduction of the growth of entanglement: Non-equilibrium}

In this section we illustrate quantitatively the effects of time-evolving the auxiliaries for the non-equilibrium setups sketched in Figure \ref{fig:setup}. We start by studying an interacting XXZ chain ($\Delta\neq0$) with two additional perturbations (dimerization $\lambda<1$ and a staggered field $b>0$) that both render the system non-integrable [see Eq.~(\ref{eq:h}) for a precise definition]. At time $t=0$, two `semi-infinite' chains (typical lengths being $L/2\sim100-200$) prepared in thermal equilibrium at temperatures $T_{L,R}$ are coupled by $h_{L/2}$ to an overall `translationally-invariant' chain. The temperature gradient drives an energy current $\langle j_{\tn{E},n}(t)\rangle$ whose typical behavior is shown in Figure \ref{fig:noneq1}(a). For $b=0$ the current at the interface $n=L/2$ saturates to a finite value on a scale $t\sim1-10$ irrespective of whether the system is gapless or gapped (indicating that either the non-integrable dimerized chain supports dissipationless transport or that its current decays on a hidden large scale; see Ref.~\onlinecite{thermalpaper} for further details). The steady-state current of the XXZ chain is of the simple `black-body' form \cite{thermalpaper,bruneau,doyon,cardyprize}
\begin{equation}
 \lim_{t\to\infty} \langle j_{\tn{E},n}(t)\rangle = f(T_L) - f(T_R)~,
\end{equation}
implying that it is determined solely by the linear conductance $G(T)\sim \partial_Tf(T)$ for arbitrary large $T_L-T_R$:
\begin{equation}
 \lim_{t\to\infty} \langle j_{\tn{E},n}(t)\rangle \sim \int_{T_L}^{T_R} G(T)\, dT~.
\end{equation}
This agrees with a field theory result \cite{doyon} at low temperatures; asymptotic low-$T$ behavior sets in around $T\lesssim0.2$.

\begin{figure*}[t]
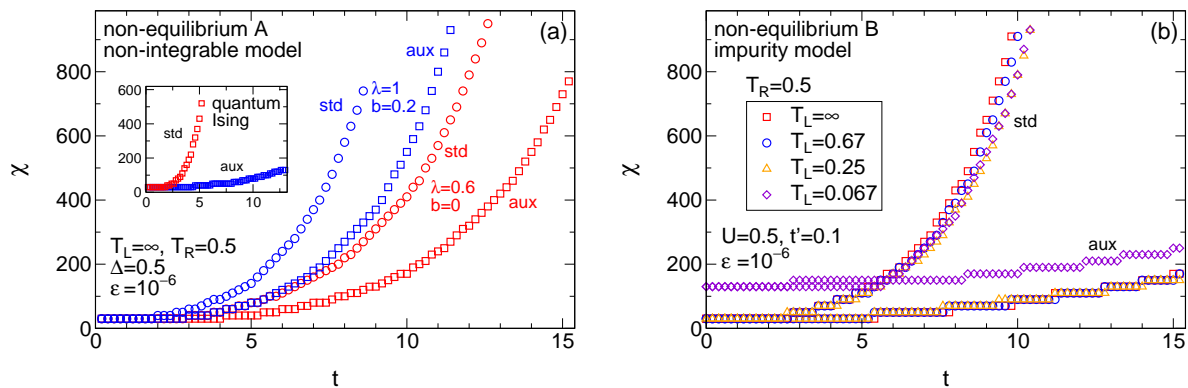

\includegraphics[width=0.48\linewidth,clip]{chi3.eps} \hspace*{0.02\linewidth}
\includegraphics[width=0.48\linewidth,clip]{chi4.eps}
\caption{(Color online) (a) The same as in Figure \ref{fig:noneq1}(b) but in presence of two perturbations (dimerization $\lambda<1$ and a staggered magnetic field $b>0$) rendering the model non-integrable. Inset: Quantum Ising model at criticality $g=1$. (b) Dimension of the MPS during the time evolution of two non-interacting XXZ chains of lengths $L/2=100$ which feature different temperatures $T_L$ and $T_R$ and are coupled at time $t=0$ via an interacting resonant level model. }
\label{fig:noneq2}
\end{figure*}

The time evolution of the corresponding bond dimension $\chi$ is shown in Figure \ref{fig:noneq1}(b) for the integrable XXZ chain both for parameters where it is gapless ($\Delta=0.5$) and gapped ($\Delta=2$). Figure \ref{fig:noneq2}(a) shows the same in presence of non-integrable perturbations as well as for the quantum Ising model of Eq.~(\ref{eq:h2}) at criticality $g=1$. In all cases and for all temperatures from $T_{L,R}=\infty$ down to $T_{L,R}=0.1$ (which for this problem corresponds to zero temperature), time evolution of the auxiliaries leads to a slower increase of $\chi$; note that the computational cost of a singular value decomposition (which dominates the whole DMRG algorithm) scales as $\chi^3$.

The same reduction of $\chi$ holds for the non-equilibrium impurity problem depicted in Figure \ref{fig:setup}(b) where an interacting resonant level model defined in Eq.~(\ref{eq:irlm}) is coupled to two non-interacting leads ($\Delta=b=0$, $\lambda=1$) at different temperatures. An exemplary evolution of the MPS dimension is shown in Figure \ref{fig:noneq2}(b); a discussion of the (transport) physics at small interactions or zero temperature can be found in Refs.~\onlinecite{irlm}. Note that due to the appearance of a Kondo-like scale, $T/\tn{bandwidth}=0.05$ does not necessarily correspond to zero temperature for this problem \cite{irlm}.

The discarded weight $\epsilon=10^{-6}$ in Figure \ref{fig:noneq1}(b) is chosen small enough to `accurately' determine the current and in particular its steady-state value. This is illustrated in Figure \ref{fig:noneq1}(a) where $\epsilon$ is successively lowered from $\epsilon=10^{-4}$ to $\epsilon=10^{-7}$. Moreover, the non-interacting case $\Delta=b=0$, $\lambda=1$ allows for an exact evaluation of the steady-state current \cite{bruneau}, which can be used to benchmark the accuracy of the DMRG data (a comparison can be found in Ref.~\onlinecite{thermalpaper}). The inset to Figure \ref{fig:noneq1}(a) shows a generic example for the \textit{actual} discarded weight (for the precise definition of $\epsilon$ see \ref{sec:dw}). It is always bound by $2\epsilon$; the total discarded weight during a complete Trotter step $e^{-iH\Delta t}e^{i\tilde H\Delta t}$ is typically bound by $10\epsilon$. It is also instructive to carry out a calculation at a fixed MPS dimension $\chi$ rather than a fixed discarded weight. The result is shown in Figure \ref{fig:noneq1}(a).

\begin{figure*}[t]
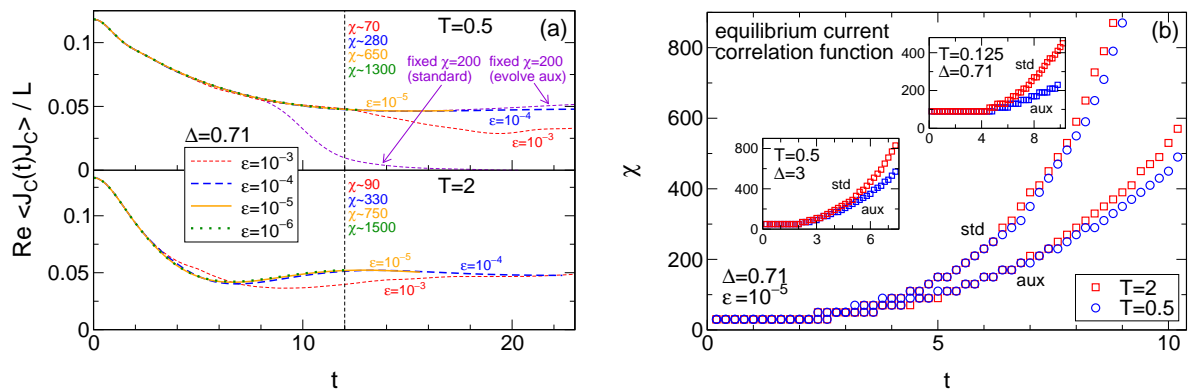

\includegraphics[width=0.48\linewidth,clip]{drude2.eps} \hspace*{0.02\linewidth}
\includegraphics[width=0.48\linewidth,clip]{chi1a.eps}
\caption{(Color online) (a) DMRG calculation of the global charge current correlation function $\langle J_\tn{C}(t)J_\tn{C}\rangle$ of the integrable XXZ chain (length $L=100$) in thermal equilibrium. The calculation is stopped once $\chi$ reaches values around $\chi\sim1500$; the numbers in the plot denote $\chi$ at time $t=12$ for the different discarded weights (the auxiliaries are time-evolved). (b) Evolution of the corresponding MPS dimension for different anisotropies $\Delta=0.71$ (gapless) and $\Delta=3$ (gapped).}
\label{fig:curr}
\end{figure*}

\subsection{Reduction of the growth of entanglement: Equilibrium}

We now turn to current correlation functions in thermal equilibrium. Following Eqs.~(\ref{eq:current}) and (\ref{eq:corr1}), we need to evaluate
\begin{equation}
e^{-iHt}U_Q(t) j_{n} |\Psi_T\rangle~.
\end{equation}
The AC conductivity $\sigma(\omega)$ is determined by the Fourier transform of $\langle J(t)J\rangle$ via Eq.~(\ref{eq:cond}). The integrable gapless XXZ chain, on which we focus in this section, supports dissipationless transport at finite temperature, i.e.~its Drude weight $D$ in Eq.~(\ref{eq:cond2}) is finite. $D$ can be obtained from the long-time limit of the current-current correlation function via Eq.~(\ref{eq:drude}). The relevant time scale in this problem is thus the scale on which $\langle J(t)J\rangle$ saturate to their asymptote.

\begin{figure*}[t]
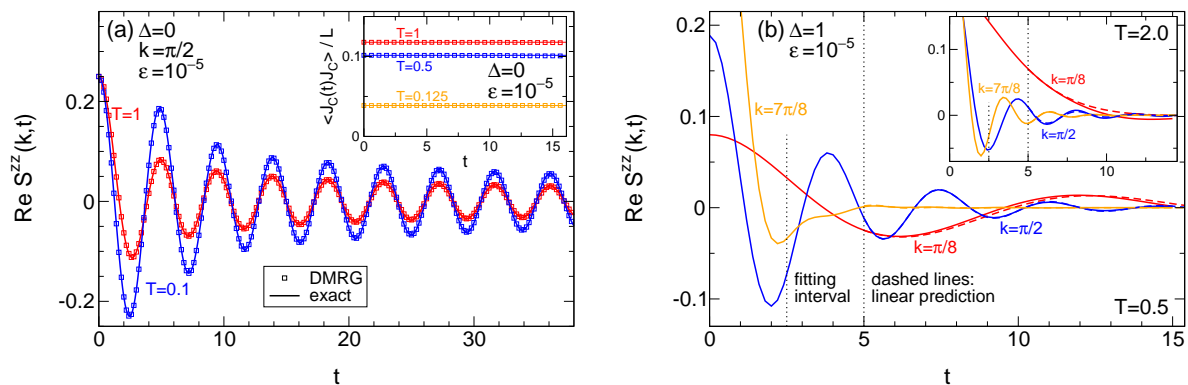

\includegraphics[width=0.48\linewidth,clip]{D0.0.eps} \hspace*{0.02\linewidth}
\includegraphics[width=0.48\linewidth,clip]{szT0.5.eps}
\caption{(Color online) (a) Equilibrium spin and global charge current correlation functions $S^{zz}(k,t) = \sum_n e^{ikn} \langle S^z_{n+L/2}(t)S^z_{L/2}\rangle$ and $\langle J_\tn{C}(t)J_\tn{C}\rangle$ of the XXZ chain at $\Delta=0$. DMRG results are compared with the exact solution obtained by mapping the model to free fermions. Since $J_\tn{C}$ is conserved, $\langle J_\tn{C}(t)J_\tn{C}\rangle$ is constant (we dropped some DMRG data points to improve visibility). However, every single term $\langle j_{\tn{C},n}(t)j_{\tn{C},m}\rangle$ that contributes to the sum is time-dependent. (b) Demonstration of the stability of linear prediction (see also Ref.~\onlinecite{barthel} and Refs.~\onlinecite{dmrgrev,trick2a,integrabilitypaper,linpred,linpred2}) for the isotropic chain $\Delta=1$. Dashed lines were obtained by fitting the DMRG data for times $t_\tn{fit}\in[2.5,5]$ and subsequent extrapolation to $t>5$ using linear prediction. The extrapolated data almost coincides with the DMRG data for $t>5$ (solid lines).}
\label{fig:sz}
\end{figure*}

DMRG data for the charge current correlation function at $\Delta=0.71$ and for various discarded weights is shown in Figure \ref{fig:curr}(a). At intermediate to high temperatures $T\gtrsim 0.5$, we can access time scales on which $\langle J(t)J\rangle$ saturates [we stop our simulation once the MPS dimension has reached values around $\chi\sim1500$; see the numbers given in Figure \ref{fig:curr}(a)]. More details can be found in Refs.~\onlinecite{drudepaper,drudepaper2}. A fairly large $\epsilon\sim10^{-5}$ is sufficient to obtain $D$ quantitatively. This is supported by the comparison with the exact solution for $\Delta=0$ shown in the inset to Figure \ref{fig:sz}(a). Note that $\langle J_\tn{C}(t)J_\tn{C}\rangle$ is constant for $\Delta=0$ since $J_\tn{C}$ is conserved; however, every single term $\langle j_{\tn{C},n}(t)j_{\tn{C},m}\rangle$ that contributes to the sum is time-dependent. It is again instructive to carry out a calculation using a constant $\chi$ rather than a fixed discarded weight; the result is shown in the upper panel of Figure \ref{fig:curr}(a).

For all temperatures from $T=\infty$ down to $T=0.125$ time evolution of the auxiliaries leads to a slower increase of the dimension of the matrix product state. This is depicted in Figure \ref{fig:curr}(b). One might expect low-$T$ behavior of $D(T)$ to set in for $T=0.125$ (see Refs.~\onlinecite{sirker1,sirker2,drudepaper2}), but the current correlation function decays so slowly that we cannot access its asymptotics without extrapolation. Exemplary data for the gapped phase is shown in the lower inset to Figure \ref{fig:curr}(b).

\subsection{Spin correlation function}

We briefly present data for the spin-spin correlation function
\begin{equation}
S^{zz}(n,t) = \langle S^z_{n+L/2}(t)S^z_{L/2}\rangle~,~~
S^{zz}(k,t) = \sum_n e^{ikn} S^{zz}(n,t)~,
\end{equation}
which provides a standard testing ground for DMRG approaches to time-dependent correlation functions at finite temperature \cite{tmrg1,tmrg2,barthel,drudepaper}. The XX chain $\Delta=0$ again allows for an exact solution by mapping the model to free fermions. Figure \ref{fig:sz}(a) shows a comparison of DMRG data with the exact result for $T=1$ and $T=0.1$ (which for this situation corresponds to low temperature). A large discarded weight $\epsilon\sim10^{-5}$ is sufficient to reproduce the analytic result up to times $t\sim30$. If the auxiliaries are time-evolved, the MPS dimension increases only moderately to $\chi\sim100-200$ in contrast to the standard approach (see Ref.~\onlinecite{drudepaper} and also Refs.~\onlinecite{tmrg1,tmrg2} for transfer-matrix DMRG data). For finite $\Delta$, the choice of $U_Q(t)=e^{i\tilde Ht}$ still outperforms $U_Q=1$ (expect at low temperatures), but the difference becomes successively less pronounced \cite{trick2b}. For the isotropic chain $\Delta=1$ and all $T\gtrsim0.1$, the accessible time scales are about a factor $1.5-2$ larger if the auxiliaries are evolved in time. In Ref.~\onlinecite{barthel} (see also Refs.~\onlinecite{dmrgrev,trick2a,integrabilitypaper,linpred,linpred2}), linear prediction was first established as an accurate tool to extrapolate correlation functions by considering exactly-solvable models and subsequently used to compute the Fourier transform of $S^{zz}(k,t)$ of the isotropic chain. For reasons of completeness, we revisit the isotropic chain and compare the result of linear prediction with DMRG data for larger times. Figure \ref{fig:sz}(b) illustrates that the agreement is good.

\begin{figure}[t]
\centering\includegraphics[width=0.5\linewidth,clip]{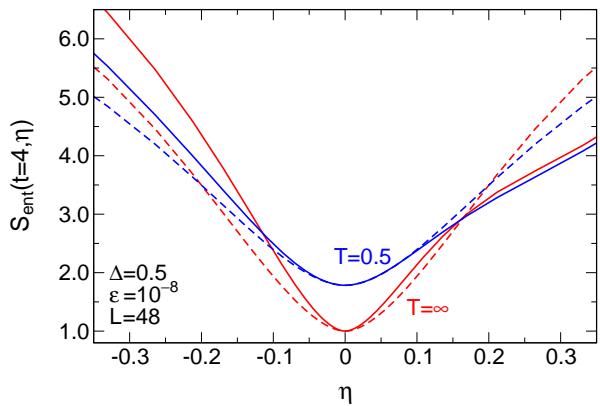}
\caption{(Color online) Entanglement entropy of the state $|\Psi_T\rangle$ after time-evolving it up to a time $t=4$. The physical degrees of freedom are time-evolved via $\exp(-iHt)$; the auxiliaries are time-evolved using $U_Q^\eta(t)=\exp(+i\tilde Ht + it\eta N)$ where $N$ is given by Eq.~(\ref{eq:N1}) (solid lines) and Eq.~(\ref{eq:N2}) (dashed lines). }
\label{fig:optimal}
\end{figure}

\subsection{Optimization of $U_Q$}

\begin{figure*}[t]
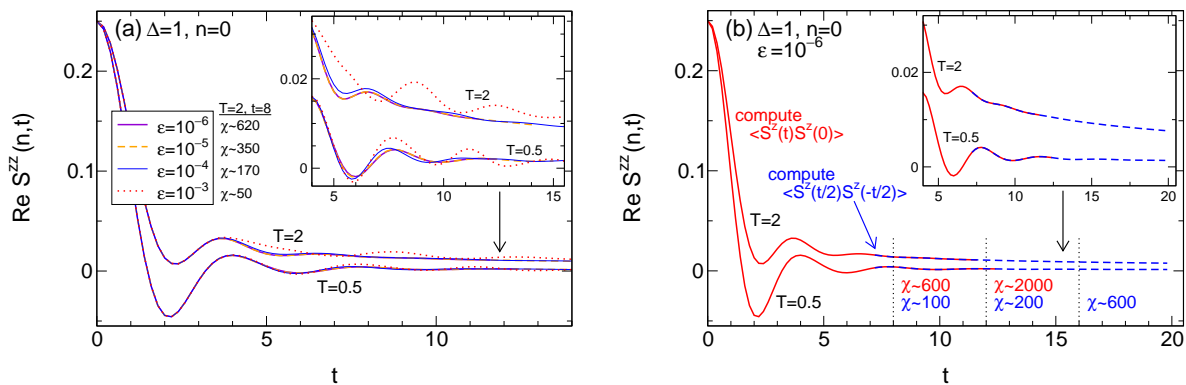

\includegraphics[width=0.48\linewidth,clip]{trick1.eps} \hspace*{0.02\linewidth}
\includegraphics[width=0.48\linewidth,clip]{trick2.eps}
\caption{(Color online) (a) Equilibrium spin correlation functions $S^{zz}(n,t)=\langle S^z_{n+L/2}(t)S^z_{L/2}\rangle$ of the isotropic XXZ chain for various discarded weights. The calculations were stopped once the MPS dimension reached values around $\chi\sim2000$. (b) By rewriting $S^{zz}(n=0,t)=\langle S^z_{L/2}(t/2)S^z_{L/2}(-t/2)\rangle$ and time-evolving $S^z_{L/2}(\pm t/2)$ independently the time scale $t$ can be accessed by a MPS of a smaller dimension $\chi$ (this idea was introduced in Refs.~\onlinecite{trick2a,trick2b}). }
\label{fig:trick}
\end{figure*}

Finally, we shortly investigate to what extend our choice of $e^{i\tilde Ht}$ for the time evolution of the auxiliaries is optimal (more details on the optimization of $U_Q$ can be found in Refs.~\onlinecite{trick2a,trick2b}). To this end, we consider
\begin{equation}
U_Q^\eta(t) = e^{+i\tilde Ht + it\eta N}~,
\end{equation}
with $N$ begin an arbitrary Hermitian matrix. We compute the entanglement entropy
\begin{equation}
S_\tn{ent}(t,\eta) =-2\sum_{a_n}(\Lambda^{L/2}_{a_n})^2\log_2\Lambda^{L/2}_{a_n} -2\sum_{a_{n_Q}}(\Lambda^{L/2_Q}_{a_{n_Q}})^2\log_2\Lambda^{L/2_Q}_{a_{n_Q}}
\end{equation}
between the left and right halves of the time-evolved state $e^{-iHt} U_Q^\eta(t) |\Psi_T\rangle$ which purifies the density matrix. We have verified that for arbitrary $N$ coupling nearest neighbors, $S_\tn{ent}(t,\eta)$ is at least quadratic in $\eta$. Figure \ref{fig:optimal} illustrates this for the two choices
\begin{equation}\label{eq:N1}
N = 1 + 2 \sum_{n_Q=1}^{L-1} \big(S^x_{n_Q}S^x_{n_Q+1} + S^y_{n_Q}S^y_{n_Q+1}\big)
\end{equation}
as well as 
\begin{equation}\label{eq:N2}
N = \sum_{n_Q=1}^{L-1}
|\sigma_{n_Q}\hspace*{-0.08cm}=\downarrow,\sigma_{n_Q+1}\hspace*{-0.08cm}=\uparrow\rangle
\langle\sigma_{n_Q}\hspace*{-0.08cm}=\downarrow,\sigma_{n_Q+1}\hspace*{-0.08cm}=\uparrow\hspace*{-0.08cm}|~.
\end{equation}
As mentioned above, Refs.~\onlinecite{trick2a,trick2b} present thorough details on how to further optimize finite-temperature dynamical DMRG. We here reiterate one of the main ideas using the language of purification. In thermal equilibrium, one can recast any correlation function exploiting time translation invariance:
\begin{equation}
\langle A(t)B\rangle = \langle A(t/2)B(-t/2)\rangle~.
\end{equation}
Thus, one only needs to carry out time evolutions up to times $|t/2|$ in order to compute $\langle A(t)B\rangle$. If $A=A^\dagger=B$ (e.g., for current-current correlation functions), it is sufficient to perform a single calculation
\begin{equation}
e^{iHt/2}U_Q^\dagger(t/2) A e^{-iHt/2}U_Q(t/2) |\Psi_T\rangle~. 
\end{equation}
This implies that one can access a time scale twice as large without much additional effort. This is illustrated in Figure \ref{fig:trick}(b) for the spin-autocorrelation function of the isotropic Heisenberg chain.

\section{Summary}

We presented extensive quantitative data for how the growth of entanglement in finite-temperature dynamical DMRG calculations can be reduced by time-evolving the auxiliary degrees of freedom which purify the thermal statistical operator. The time evolution of the auxiliaries is an exact modification to the standard algorithm. Our particular focus was on energy and charge transport properties both in linear response (described by current-current correlation functions) and in non-equilibrium driven by temperature gradients. We studied a variety of gapless and gapped integrable and non-integrable spin-$1/2$ chains (i.e., interacting spinless fermions), the quantum Ising model at criticality, and impurity (quantum dot) setups. For all temperatures from $T=\infty$ down to $T/\tn{bandwidth}=0.05$, which for most problems investigated in this work corresponds to low temperatures, the build-up of entanglement is reduced. This speeds up numerics and allows to eventually access longer time scales.

\vspace*{1ex}

\emph{Acknowledgments} --- We are grateful to Frank Verstraete for very useful suggestions and thank Thomas Barthel, Fabian Heidrich-Meisner, and Steve White for discussions. Support by the Deutsche Forschungsgemeinschaft via KA3360-1/1 (CK), by the DARPA TI program of UCLA (JHB) as well as by the Nanostructured Thermoelectrics program of LBNL (CK and JEM) is acknowledged.

\appendix

\section{Technical details.}
\label{sec:appendix}

\subsection{Purification.}

One way to evaluate Eqs.~(\ref{eq:corr}) and (\ref{eq:rhot}) within the density matrix renormalization group is to purify \cite{purification} the thermal density matrix by introducing an auxiliary Hilbert space $Q$:
\begin{equation}
\rho_T = \frac{e^{-H/T}}{Z_T} = \tn{Tr\,}_{Q} |\Psi_T\rangle\langle\Psi_T|~.
\end{equation}
At infinite temperature, where $\rho_T$ is proportional to the unit operator, one can analytically write down the purification:
\begin{equation}
\rho_\infty =\frac{1}{Z_\infty}=\frac{1}{2^{L}} = \tn{Tr}_{Q}|\Psi_\infty\rangle\langle\Psi_\infty|  = \tn{Tr}_{Q}\prod_{n=1}^{L} |\Psi_\infty^{n}\rangle\langle\Psi_\infty^{n}|~.
\end{equation}
In order to exploit conservation laws within the DMRG algorithm, it is convenient to choose
\begin{equation}\label{eq:initialstate}
|\Psi_\infty^{n}\rangle = \frac{1}{\sqrt{2}}\left( \left|\sigma_n\hspace*{-0.08cm}=\uparrow,\sigma_{n_Q}\hspace*{-0.08cm}=\downarrow\right\rangle - \left|\sigma_n\hspace*{-0.08cm}=\downarrow,\sigma_{n_Q}\hspace*{-0.08cm}=\uparrow\right\rangle \right)~,
\end{equation}
where $\sigma_n\hspace*{-0.08cm}=\uparrow,\downarrow$ denotes the eigenbasis of $S^z_n$, and $Q$ is spanned by
\begin{equation}
Q = \tn{span}\, \big\{ |\sigma_{1_Q}\ldots\sigma_{L_Q}\rangle \big\} ~.
\end{equation}
One readily verifies that indeed
\begin{equation}
\frac{1}{2} = \frac{1}{2}\sum_{\sigma_{n}} |\sigma_{n}\rangle\langle\sigma_{n}| =
\sum_{\sigma_{n}}\sum_{\sigma_{n_Q}}\langle\sigma_{n_Q}|\Psi_\infty^{n}\rangle\langle\Psi_\infty^{n}|\sigma_{n_Q}\rangle\,.
\end{equation}
The finite-temperature purified state $|\Psi_T\rangle$ is obtained from $|\Psi_\infty\rangle$ by applying an imaginary time evolution \cite{dmrgrev},
\begin{equation}\begin{split}
 \rho_T & = \frac{e^{-H/T}}{Z_T} = \frac{Z_\infty}{Z_T}\tn{Tr}_Q\, e^{-H/2T} |\Psi_\infty\rangle\langle\Psi_\infty|e^{-H/2T}\\
& =  \frac{1}{\langle \Psi_T|\Psi_T\rangle}\tn{Tr}_Q |\Psi_T\rangle\langle\Psi_T|~,~~~|\Psi_T\rangle=e^{-H/2T}|\Psi_\infty\rangle ~,
\end{split}\end{equation}
where we formally replaced
\begin{equation}\label{eq:hq}
H \to H\otimes 1_Q~.
\end{equation}
The correlation function of Eq.~(\ref{eq:corr}) is \textit{exactly} recast as
\begin{equation}\label{eq:corr1}
\langle A(t)B\rangle = \frac{\langle \Psi_\infty| e^{-\frac{H}{2T}}e^{iHt} A e^{-iHt} B e^{-\frac{H}{2T}}|\Psi_\infty\rangle}{\langle\Psi_\infty| e^{-\frac{H}{T}}|\Psi_\infty\rangle}~,
\end{equation}
and for two time arguments one similarly finds
\begin{equation}
\langle A(t)B(t')\rangle = \frac{\langle \Psi_T| e^{iHt} A e^{-iHt}e^{iHt'} Be^{-iHt'}|\Psi_T\rangle}{\langle\Psi_T|\Psi_T\rangle}~. 
\end{equation}
In order to evaluate Eq.~(\ref{eq:intro2}) for $A=A^\dagger=B$, it is sufficient to compute
\begin{equation}\label{eq:corr2}
e^{iHt/2} A e^{-iHt/2} |\Psi_T\rangle~. 
\end{equation}
The expectation value of any observable $A$ with respect to the time-dependent (non-equilibrium) density matrix of Eq.~(\ref{eq:rhot}) is given by
\begin{equation}\label{eq:corr3}
\langle A(t) \rangle = \tn{Tr}\,\rho(t) A
= \frac{\langle \Psi_\infty| e^{-\frac{H_L}{2T_L}-\frac{H_R}{2T_R}}e^{iHt} A e^{-iHt}e^{-\frac{H_L}{2T_L}-\frac{H_R}{2T_R}}|\Psi_\infty\rangle}{\langle\Psi_\infty| e^{-\frac{H_L}{2T_L}-\frac{H_R}{2T_R}}|\Psi_\infty\rangle}~.
\end{equation}

\subsection{DMRG algorithm}

\subsubsection{Initial state.}

It is convenient to first express the initial state $|\Psi_\infty\rangle$ in terms of a matrix product state \cite{mps1,mps2,verstraete1,verstraete2},
\begin{equation}\label{eq:mps}
|\Psi_\infty\rangle = \sum_{\sigma_n,\sigma_{n_Q}} A^{\sigma_1}A^{\sigma_{1_Q}}\cdots A^{\sigma_{L}}A^{\sigma_{L_Q}}|\sigma_1\sigma_{1_Q}\ldots\sigma_{L}\sigma_{L_Q}\rangle~,
\end{equation}
where
\begin{equation}
A^{\sigma_i}_{a_ia_{i+1}} = \Lambda_{a_i}^i\Gamma_{a_ia_{i+1}}^{\sigma_i}~.
\end{equation}
Here and in the following $\sigma_i$ is a short hand for either a physical or auxiliary degrees of freedom: $\sigma_i\in\{\sigma_n,\sigma_{n_Q}\}$. The initial matrices corresponding to Eq.~(\ref{eq:initialstate}) read
\begin{align}
\Gamma^{\sigma_{n_{\phantom{Q}}}=\uparrow}& = (1~~0) &
\Gamma^{\sigma_{n_{\phantom{Q}}}=\downarrow}& = (0~~-1) \nonumber \\
\Gamma^{\sigma_{n_Q}=\uparrow}& = (0~~1/\sqrt{2})^T &
\Gamma^{\sigma_{n_Q}=\downarrow}& = (1/\sqrt{2}~~0)^T ~,
\end{align}
as well as $\Lambda^i=1$. Normalization generally stipulates
\begin{equation}\label{eq:norm1}
\sum_{\sigma_ia_i} (\Lambda^i_{a_i}\Gamma^{\sigma_i}_{a_ia_{i+1}})^\dagger (\Lambda_{a_i}^i\Gamma^{\sigma_i}_{a_ia_{i+1}'}) = \delta_{a_{i+1}a_{i+1}'}
\end{equation}
as well as
\begin{equation}\label{eq:norm2}
\sum_{\sigma_ia_{i+1}} (\Gamma^{\sigma_i}_{a_ia_{i+1}}\Lambda^{i+1}_{a_{i+1}})(\Gamma^{\sigma_i}_{a_i'a_{i+1}}\Lambda_{a_{i+1}}^{i+1})^\dagger = \delta_{a_{i}a_{i}'}~.
\end{equation}

\subsubsection{Trotter decomposition.}

In order to successively apply the imaginary and real time evolutions operators
\begin{equation}
e^{-\lambda H} = e^{-\Delta\lambda H}e^{-\Delta\lambda H}\ldots ~~,~~~~\lambda = \Delta\lambda+ \Delta\lambda + \ldots
\end{equation}
to the initial state $|\Psi_\infty\rangle$, we factorize them either via a second or a fourth order Trotter decomposition \cite{dmrgrev}. The second order scheme reads
\begin{equation}
e^{-\Delta\lambda H} \approx e^{-\Delta\lambda/2 H_1} e^{-\Delta\lambda H_2} e^{-\Delta\lambda/2 H_1}~,
\end{equation}
where the $H=H_1+H_2$, and $H_{1,2}$ contain only terms which mutually commute:
\begin{equation}
H_1 = \sum_{n\tn{ odd}} h_n~,~~H_2 = \sum_{n\tn{ even}} h_n~.
\end{equation}
In order to reduce the error, one can employ a fourth order decomposition \cite{suzuki},
\begin{equation}
e^{-\Delta\lambda H} \approx U(\Delta\lambda_1)^2\, U(\Delta\lambda_2)\, U(\Delta\lambda_1)^2~,
\end{equation}
where
\begin{equation}
U(\Delta\lambda_i) = e^{-\Delta\lambda_i H_1/2}e^{-\Delta\lambda_i H_2}e^{-\Delta\lambda_i H_1/2}~,
\end{equation}
and
\begin{equation}
\Delta\lambda_1=\frac{1}{4-4^{1/3}}\Delta\lambda~,~~\Delta\lambda_2=\Delta\lambda-4\Delta\lambda_1~.
\end{equation}
We typically employ a second order Trotter decomposition with a time step of $\Delta\beta=0.005$ during the imaginary time evolution from $\beta=0$ down to the physical temperature $\beta=1/T$ and a fourth order decomposition with a time step of $\Delta t=0.2$ during the real time evolution. This is sufficient for all problems studied in this paper (which we verify by comparing against data obtained for smaller $\Delta\beta=0.0025$ and $\Delta t=0.1$). The Trotter decomposition is typically not a significant source of error within time-dependent DMRG calculations \cite{noneqtherm1}.

\subsubsection{DMRG step.}

The physical Hamiltonians of Eqs.~(\ref{eq:h}) and (\ref{eq:h2}) couple matrices with indices $\sigma_n$ and $\sigma_{n+1}$ which in the matrix product state of Eq.~(\ref{eq:mps}) are next-nearest neighbors -- they are separated by a matrix associated with an auxiliary degree of freedom $\sigma_{n_Q}$. Through Eq.~(\ref{eq:hq}), the local Hamiltonians of Eqs.~(\ref{eq:h}) and (\ref{eq:h2}) are formally replaced by
\begin{equation}
h_n \to h_n\otimes \sum_{\sigma_{n_Q}}|\sigma_{n_Q}\rangle\langle\sigma_{n_Q}|~.
\end{equation}
Thus, after each application of $\exp(-\Delta\lambda h_n)$, two singular value decompositions are carried out to update three consecutive matrices; to keep the notation transparent, let us denote those matrices as:
\begin{align}
\Gamma^{\sigma_{0} } & \longrightarrow \tilde \Gamma^{\sigma_{0}} & & \nonumber \\
\Gamma^{\sigma_{1}} & \longrightarrow \tilde \Gamma^{\sigma_{1}} & & 
\Lambda^{1}\longrightarrow \tilde \Lambda^{1}\nonumber \\
\Gamma^{\sigma_{2}} & \longrightarrow \tilde \Gamma^{\sigma_{2}} & & 
\Lambda^{2}\longrightarrow \tilde \Lambda^{2}~,
\end{align}
and furthermore abbreviate $h=h_n$. This `DMRG update' can be achieved by a simple and straightforward generalization of the protocol for nearest neighbors outlined in Ref.~\onlinecite{dmrgrev}. First, one forms the three-site tensor,
\begin{equation}
\Psi_{a_0a_3}^{\sigma_0\sigma_1\sigma_2} = \sum_{a_1a_2}
\Lambda_{a_0}^0\Gamma^{\sigma_0}_{a_0a_1}
\Lambda_{a_1}^1\Gamma^{\sigma_1}_{a_1a_2}
\Lambda_{a_2}^2\Gamma^{\sigma_2}_{a_2a_3}
\Lambda_{a_3}^3~,
\end{equation}
which is then acted on by $\exp(-\Delta\lambda h)$:
\begin{equation}
\Phi_{a_0a_3}^{\sigma_0\sigma_1\sigma_2} =
\sum_{\sigma_1'\sigma_2'\sigma_3'} \Psi_{a_0a_3}^{\sigma_0'\sigma_1'\sigma_2'}
e^{-\Delta\lambda h}\big|_{\sigma_0\sigma_1\sigma_2}^{\sigma_0'\sigma_1'\sigma_2'}~.
\end{equation}
Now, two singular value decompositions (SVDs) are applied to appropriately reshaped tensors in order to obtain the new matrices $\tilde\lambda^i$ and $\tilde\Gamma^{\sigma_i}$:
\begin{equation}\label{eq:svd1}\begin{split}
\Phi_{a_0a_3}^{\sigma_0\sigma_1\sigma_2} & =
\Phi_{(a_0\sigma_0),(a_3\sigma_1\sigma_2)} \\
& = \sum_{a_1} U_{(a_0\sigma_0),a_1} \tilde\Lambda^1_{a_1} (V^\dagger)_{a_1,(\sigma_1\sigma_2a_3)}\\
& = \sum_{a_1}\Lambda^0_{a_0}\underbrace{\Lambda^{0,-1}_{a_0}U^{\sigma_0}_{a_0a_1}}_{\tilde\Gamma^{\sigma_0}_{a_0a_1}}
\underbrace{\tilde\Lambda^1_{a_1} (V^\dagger)_{a_1,(\sigma_1\sigma_2a_3)}}_{\tilde\Phi_{a_1,(\sigma_1\sigma_2a_3)}}~,
\end{split}\end{equation}
and likewise
\begin{equation}\label{eq:svd2}\begin{split}
\tilde\Phi_{a_1,(\sigma_1\sigma_2a_3)} & = \tilde\Phi_{(a_1\sigma_1),(\sigma_2a_3)} \\
& = \sum_{a_2} U_{(a_1\sigma_1),a_2}\tilde\Lambda_{a_2}^2(\tilde V^\dagger)_{a_2,(\sigma_2a_3)} \\
& = \sum_{a_2} \tilde\Lambda_{a_1}^1 \underbrace{\tilde\Lambda^{1,-1}_{a_1}U^{\sigma_1}_{a_1a_2}}_{\tilde\Gamma^{\sigma_1}_{a_1a_2}}
\tilde\Lambda_{a_2}^2 \underbrace{(\tilde V^\dagger)_{a_2a_3}^{\sigma_2} \Lambda_{a_3}^{3,-1}}_{\tilde\Gamma^{\sigma_2}_{a_2a_3}}\Lambda_{a_3}^3~.
\end{split}\end{equation}
In case of a real time evolution, the normalization conditions of Eqs.~(\ref{eq:norm1}) and (\ref{eq:norm2}) are preserved automatically (up to errors associated with a potential truncation of the matrices; see below). For a non-unitary $\exp(-\Delta\beta h)$, however, the matrix product state needs to be brought back to its canonical form in order to simplify the evaluation of expectation values, and, more importantly, because otherwise the truncation below is not optimal. Normalization can be achieved approximately by successively acting with unit operators $\Delta\lambda=0$ \cite{tebd} or exactly using the procedure outlined in Ref.~\onlinecite{norm1}.

\subsubsection{The discarded weight.}
\label{sec:dw}

Up to this point, the presented DMRG algorithm to evaluate Eqs.~(\ref{eq:corr1})--(\ref{eq:corr3}) is exact (except for the Trotter error which can be made sufficiently small for all problems studied in this work). However, the dimension of the matrices in Eq.~(\ref{eq:mps}) increases during each of the singular value decompositions in Eqs.~(\ref{eq:svd1}) and (\ref{eq:svd2}), generically by a factor of two (corresponding to the number of local degrees of freedom). For brevity, let us neglect that for open boundary conditions the matrix dimension is position-dependent and simply denote it by $\chi$ (a more thorough discussion can be found in Ref.~\onlinecite{dmrgrev}). The computational effort of the DMRG algorithm is dominated by the cost of the singular value decompositions, which scales as $\chi^3$. Increasing $\chi\to2\chi$ during each sub-step of a Trotter step is numerically unfeasible (and fortunately generically unnecessary). Thus, we truncate the matrices to a dimension $\chi'<2\chi$ by dropping the indices $a_i$ which correspond to the smallest singular values $\tilde\Lambda_{a_i}$. For a normalized state [i.e., if Eqs.~(\ref{eq:norm1}) and (\ref{eq:norm2}) hold], this is the best approximation with respect to the Hilbert space norm. The error (i.e., the norm-distance of the states before and after truncation) is given by the sum of all squared singular values which are discarded during all sub-steps of one Trotter step. This so-called discarded weight is the key numerical control parameter; the DMRG algorithm becomes exact as it is sent to zero.

One can strictly enforce a fixed discarded weight by analyzing the singular values during all SVDs carried out within one Trotter substep and then retrospectively choosing $\chi'$ appropriately (or by simply repeating the substep if $\chi'>\chi$). More pragmatically, one can fix $\chi'=\chi$ and increase $\chi$ by $\Delta\chi$ after one substep if the total discarded weight exceeds a pre-defined threshold $\epsilon$. While this simple approach certainly becomes inapplicable if one aims at strictly maintaining a discarded weight that is very small, it works for all problems studied in this paper, with $\Delta\chi\sim10-20$ being a reasonable choice. We observe that the actual discarded weight is smaller than $2\epsilon$ for all data presented in this paper [it is imperative to always monitor the real discarded weight which determines the actual error; this is illustrated in the inset to Figure \ref{fig:noneq1}(a)]. In the following we neglect this difference and simply refer to our control parameter $\epsilon$ as `discarded weight'. We successively lower $\epsilon$ from $10^{-8}-10^{-13}$ during the imaginary time evolution and from $10^{-3}-10^{-7}$ during the real time evolution, which turns out to be sufficient to obtain `accurate' results for all physical quantities of interest. `Accurate' means `sufficiently converged with respect to lowering $\epsilon$', or `in reasonable agreement with exact data' in case that the problem allows for an analytic solution; all this is illustrated in Figures \ref{fig:noneq1}(a), \ref{fig:curr}(a), \ref{fig:sz}(a), and \ref{fig:trick}(a). We stop our simulations once the matrix dimension has reached values around $\chi\sim1500-2500$.

\vspace*{3ex}



\begin{thebibliography}{*}
\bibitem{commentconteq}   
In the charge case we focus on $\lambda=1$.
\end{thebibliography}
\end{document}